\documentclass[modern]{aastex61}
\usepackage{natbib}
\usepackage{graphicx}
\usepackage{subfigure}
\usepackage{amsmath,amssymb}
\usepackage{mathrsfs}
\usepackage{verbatim}
\begin{document}

\title{The Neupert Effect of Flare UltraViolet and Soft X-ray Emissions}
%
%
%
\author{Jiong Qiu}
\affil{Department of Physics, Montana State University, Bozeman, MT, USA}

\begin{abstract}
We model the Neupert effect that relates flare heating energy with the observed
SXR emission. The traditional form of the Neupert effect refers to the correlation
between the time-integrated HXR or microwave light curve and the SXR light curve.
In this paper, instead, we use as the proxy for heating energy
the ultraviolet (UV) emission at the foot-points of flare loops, and modify the model of the Neupert
effect by taking into account the discrete nature of flare heating as well as cooling.
In the modified empirical model, spatially resolved UV lightcurves from the transition region or
upper chromosphere are each convolved 
with a kernel function characterizing decay of the flare loop emission. Contributions 
by all loops are summed to compare with the observed total SXR emission. The model
has successfully reproduced the observed SXR emission from its rise to decay. 
To estimate heating energies in flare loops, we also employ the UV Foot-point 
Calorimeter (UFC) method that infers heating rates in flare loops from these UV light 
curves and models evolution of flare loops with a zero-dimensional hydrodynamic code. 
The experiments show that a multitude of impulsive heating events do not well reproduce 
the observed flare SXR light curve, but a two-phase heating model
leads to better agreement with observations. Comparison of the
two models of the Neupert effect further allows us to calibrate the UFC method, and 
improve the estimate of heating rates in flare loops continuously formed by 
magnetic reconnection throughout the flare evolution.

\end{abstract}

\section{INTRODUCTION}
\citet{Neupert1968} discovered that the time integral of the microwave light curve of 
a flare is correlated with the flare soft X-ray (SXR) light curve during its rise. 
Subsequently, the Neupert effect has been confirmed in generations of 
flare observations. \citet{Dennis1993} studied 66 flares observed in 1980 
by the {\it Hard X-ray Burst Spectrometer} (HXRBS) on the {\it Solar Maximum Mission} \citep[SMM;][]{Orwig1980}
and the {\it Geostationary Operational Environmental Satellite} (GOES), 
finding that 80\% of large flares exhibit good correlations 
between the hard X-ray (HXR) light curve and the time derivative of the 
GOES SXR light curve in the 1-8~\AA\ passband. Applying the time-correlation 
analysis to more than one thousand flares observed between 1997 January and 2000 June
by GOES and the {\it Burst and Transient Source Experiment} (BATSE) on-board the {\it Compton Gamma-Ray Observatory}
\citep{Schwartz1992}, \citet{Veronig2002} confirmed that the timing behaviour of the HXR and SXR 
emissions in large flares is consistent with the Neupert effect.
\citet{McTiernan1999} examined flare SXR and HXR observations by the {\it Soft X-ray Telescope}
\citep[SXT; ][]{Tsuneta1991}, the {\it Bragg Crystal Spectrometer}\citep[BCS;][]{Culhane1991}, 
and the {\it Hard X-ray Telescope} \citep[HXT;][]{Kosugi1991} on Yohkoh, finding 
the Neupert effect more prominently demonstrated in high temperature SXR light curves.
\citet{Effenberger2017} further confirmed the Neupert
effect exploiting flare observations by the {\it Reuven Ramaty High Energy Solar Spectroscopic Imager} \citep[RHESSI; ][]{Lin2002}
for the past two solar cycles. The Neupert effect has also been found in small flares. 
\citet{Qiu2004} studied the Neupert effect in more than 100 microflares 
(of GOES class A to C1) with significant HXR emissions observed
by RHESSI, finding that the time derivative of the GOES SXR 
emission is best correlated with the HXR emission at the photon energy 14 -- 20 keV. 
\citet{Glesener2020} have recently detected non-thermal HXR emission in a A5.7 microflare
observed by the {\it Nuclear Spectroscopic Telescope Array}
\citep[NuSTAR;][]{Grefenstette2016}, which also exhibits the Neupert effect.

The Neupert effect is interpreted as that flare plasmas in the corona are heated by non-thermal electrons.
These electrons precipitate at the lower atmosphere, and lose their energy instantaneously by collision
with ions. In this course, thick-target HXR emissions are generated, and chromosphere
evaporation is driven that heats the corona as well as increases the density of the corona, leading 
to the enhanced SXR emission \citep[e.g.,][]{Antonucci1982, Fisher1985b, Li1993, Lee1995}. Therefore, the
HXR light curve of a flare can serve as the proxy of the electron energy flux, and its time integral 
is equivalent to the maximum thermal energy of the subsequently heated flare plasmas in the corona, 
achieved at the time when the flare SXR emission peaks. Analyzing spectroscopic observations of flares, 
a number of studies have then estimated this maximum flare thermal energy as well as the total 
energy in non-thermal electrons, suggesting that these two energies are indeed comparable in 
large flares \citep[see ][ and references therein]{Emslie2012, Aschwanden2017c},
and sometimes in small flares as well \citep[e.g.,][]{Glesener2020}.
With this notion, generations of hydrodynamic models have been developed to study evolution
of flare corona with non-thermal electron beams as the primary source of heating \citep{Somov1981, Nagai1984,
Mariska1989, Emslie1992, Warren2004, Reep2014}. Specifically, effort has been made to model evolution
of the flare corona (and chromosphere), using observed HXR light curves or the time derivative of
SXR light curves to infer time-dependent heating rates in flares, and reproduce observed thermodynamic properties
of flare plasmas in the corona and the chromosphere \citep{Fisher1990, Rubio2016}.

Despite the prevailing evidence in support of the Neupert effect, there are several caveats in the traditional form
of the Neupert effect. It only addresses the rise phase of the flare SXR emission and only considers non-thermal 
electrons as the primary carrier of corona heating energy. As has been noted for decades, energy 
release and flare heating often continue into the decay phase of the flare SXR 
emission, when the HXR emission has usually diminished, and the amount of heating
energy deposited in the decay phase can be significant \citep{Withbroe1978, Dere1979, Ryan2013}. 
Whereas prior studies have confirmed the Neupert effect in a large number of flares, these same studies
have also revealed that, in a significant fraction of flares, the SXR emission 
continues to rise after the HXR emission has ended \citep{Veronig2002}, and 
in some flares, the SXR emission rises before the HXR emission \citep{Effenberger2017}. 
These observations indicate that other sources of energy are needed to heat the flare 
corona \citep{Veronig2005}. Furthermore, flare heating takes place in many flare loops that 
are generated continuously into the decay phase. These loops are heated, by chromosphere evaporation 
driven by either non-thermal beams or else, such as thermal conduction \citep{Gan1991, Longcope2014} or
Alfv{\'e}n waves \citep{Fletcher2008, Kerr2016}, and then cool, and the total SXR emission 
at any given time is the sum of the emissions from all these 
loops at their different evolution stages \citep[e.g.,][]{Aschwanden2001}. The continuous heating 
and cooling of multiple flare loops cannot be well described by the Neupert effect applied to the 
total HXR and SXR emissions that are not spatially resolved. 

These questions motivate the thinking to extend the Neupert effect to a broader context
that addresses the nature of flare heating on elementary scales and perhaps beyond non-thermal
electrons. Apart from microwave and HXR light curves, which are indicative of non-thermal electrons,
flare emission in the lower-atmosphere observed in the optical, ultraviolet, and extreme 
ultraviolet wavelengths generally exhibits an impulsive behavior 
before the more gradual rise of the SXR emission \citep[see the review by ][]{Fletcher2011}. 
In large flares, enhanced UV and EUV emissions have often been found to trace
HXR emissions temporally and/or spatially \citep{Kane1971, McClymont1986, Cheng1988, Cheng1990, 
Fletcher2001, Warren2001, Qiu2010, Cheng2012}, supporting the scenario of heating by
non-thermal electrons. But observations have also shown impulsive UV emissions at the
flare foot-points not associated
with thick-target HXR signatures \citep{Warren2001, Alexander2006, Coyner2009, Cheng2012},
and in these cases, it is likely that the temperature of the corona is rapidly raised,
and thermal conduction would deposite energy at the chromosphere, causing enhanced optical, 
UV, and EUV emissions, and driving chromosphere evaporation as well. Most recently,
spectroscopic observations in these wavelengths with high spatial 
resolutions have revealed downflows (chromosphere condensation) and upflows (chromosphere evaporation) 
in a large number of flare kernels at unprecedented small scales, illustrative of prototypical, elementary energy release
events in the flare \citep{Graham2015}. These state-of-the-art observations clearly demonstrate the
critical role of chromosphere evaporation in energizing the flare corona regardless of heating mechanisms.

The advanced flare observations in the lower atmosphere provide us with the opportunity to better characterize heating rates in flare loops. In this
spirit, we analyze the ultraviolet emission from the transition region and upper chromosphere
at the foot-points of flare loops. The transition region and upper chromosphere respond promptly 
to energy release in the corona, and the resultant UV emission can be used as a proxy for heating. This approach
is free from the assumption that heating is primarily by non-thermal electrons. Furthermore, high-resolution UV 
images allow us to track flare loops that are formed and heated at different times and evolve independently 
throughout the flare, assuming that these loops are anchored at brightened UV pixels. 
This paper presents a thought experiment on the Neupert effect using spatially resolved UV light curves instead
of HXR light curves, and with two models, a modified empirical model of the Neupert effect, and the UV Footpoint Calorimeter
(UFC) method that infers heating rates from UV light curves and models
evolution of the flare corona in a multitude of loops \citep{Qiu2012, Liu2013}. 
Both models take into account heating as well as cooling of 
flare loops formed at different times during the flare, which contribute to the observed total SXR emission. 
The first model examines the temporal relationship between the SXR and spatially resolved
UV 1600~\AA\ light curves but cannot return the heating energy, whereas the UFC method will be able to infer the heating rates in flare loops.
In this study, we analyze 16 flares observed by GOES and the {\em Atmospheric Imaging Assembly} 
\citep[AIA;][]{Lemen2012} (Section 2), apply the empirical model (Section 3) and UFC method (Section 4)
to these flares to reproduce the GOES SXR light curves, and improve the estimate of flare heating energies by comparing
these two models (Section 5). Conclusions and discussions are given in the last section.

\section{FLARE LIGHT CURVES}

We have analyzed 16 flares listed in Table~\ref{flareinfo}. The flare SXR emissions were obtained by 
GOES\footnote{In the table, the magnitude of the flare is based on the GOES flux in the 1 -- 8~\AA\ passband, which has been,
historically, scaled to match the flux by GOES satellites 1 -- 7. As of October 28, 2020, the SXR flux obtained
by GOES satellites 8 -- 15 is reported as the ``true" flux, which is equivalent to the ``scaled" flux divided by 0.7 
for the long channel (1-8~\AA) and by 0.85 for the short channel (0.5 -- 4~\AA), respectively (\texttt{https://hesperia.gsfc.nasa.gov/rhessidatacenter/complementary\_data/goes.html}). 
The flares analyzed in this 
paper were observed by GOES satellites 10 -- 15, and analysis in this paper uses the ``true" flux in units of W m$^{-2}$; yet to be consistent with
the past literature, the flare magnitude reported in Table~\ref{flareinfo} is still derived using the ``scaled" flux.}, 
and imaging observations of the flares in the UV 1600\AA\ passband were obtained by AIA
on board the {\it Solar Dynamics Observatory} \citep[SDO;][]{Pesnell2012}. Except for one event SOL2011-12-26 
(\#3), these flares were also observed by RHESSI. Table~\ref{flareinfo} presents the information of the source region
and position of each flare, the duration of the flare $\tau_d$ derived from the flare light curves, and the median half-length
of flare loops estimated from the separation of the flare ribbons observed in the AIA 1600\AA\ images. The magnetic flux enclosed
in the total area of the flare ribbons gives the measurement of the total reconnection flux $\Phi_{rec}$ \citep[e.g. ][]{Qiu2004a, Saba2006}, 
and the uncertainty in $\Phi_{rec}$ is characterized by the difference in the magnetic flux measured in positive and negative
magnetic fields, respectively. The total heating energy and its uncertainty in each flare are derived in the following text (Sections 3, 4 and 5.1).

\begin{figure}
\epsscale{1.2}
\plotone{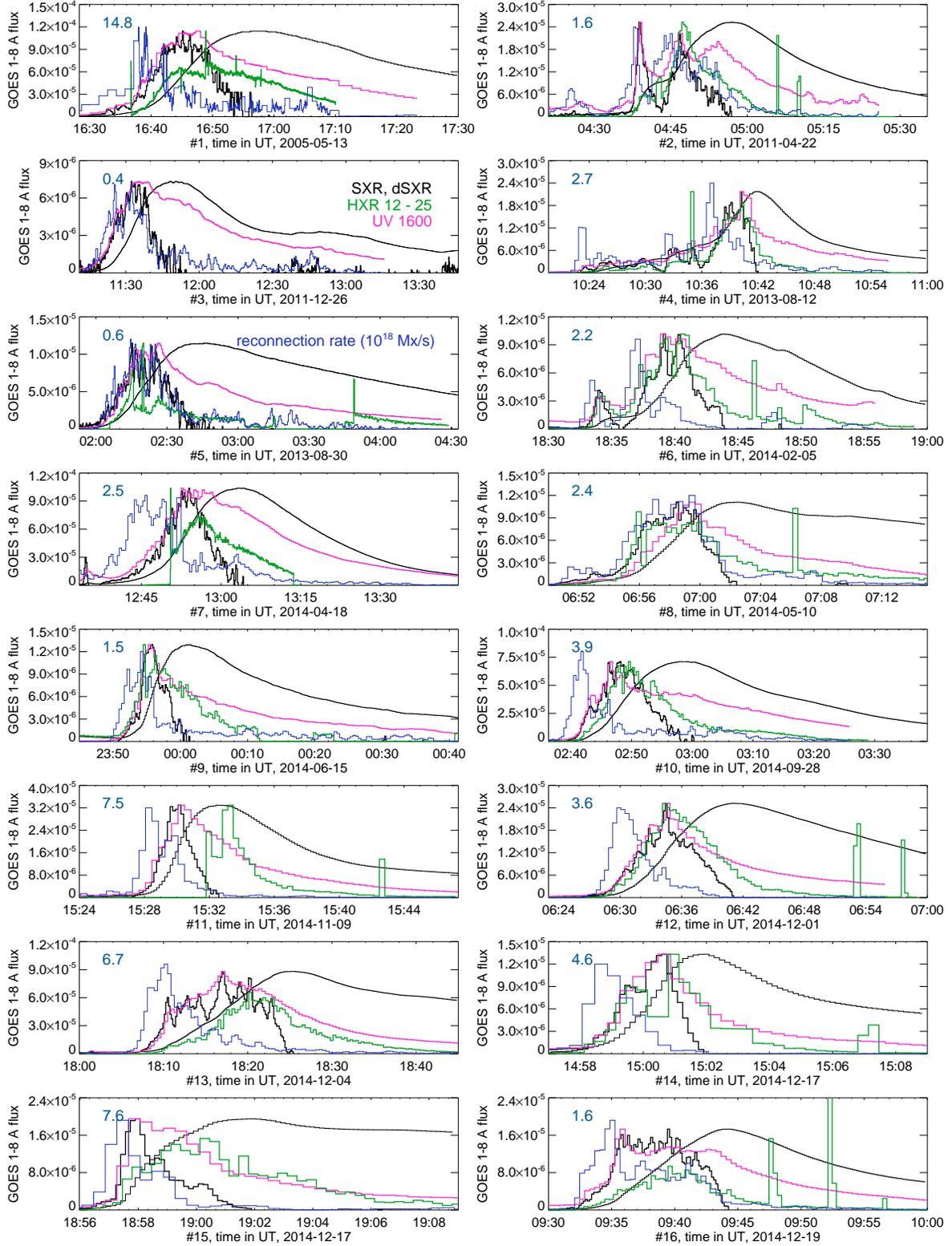}
\caption{Light curves of the flares analyzed and modeled in this paper. These include
the GOES SXR ``true" flux in 1 -- 8\AA\ in units of W m$^{-2}$ and its time derivative
(black), the total UV counts rate light curve (pink), integrated over the flare region, 
in 1600\AA\ passband from AIA/SDO, and the HXR counts rate light curve (green) at the photon 
energy 12 - 25~keV observed by RHESSI. Also plotted is
the time profile of the reconnection rate in units of 10$^{18}$ Mx s$^{-1}$ (blue),
with the peak reconnection rate marked in each panel. 
Except the SXR light curve, all other light curves are arbitrarily scaled.
For clarity of the display, the
uncertainties in the reconnection rates are not plotted, but they are described in the text (Section 5.2).
}\label{fig:lightcurve}
\end{figure}

Figure~\ref{fig:lightcurve} shows the light curves of each of the 16 flares, including the
GOES SXR light curve at 1-8~\AA\ (denoted as $\mathcal{F}_{{\rm sxr}}$, in units of W m$^{-2}$, in the following text), 
its time derivative ($\dot{\mathcal{F}}_{{\rm sxr}}$), the total counts rate light curve in the 
UV 1600\AA\ passband integrated over the flare region ($\mathcal{F}_{{\rm uv}}$, in units of DN s$^{-1}$),
and the HXR counts rate light curve of photon energy 12 - 25~keV by RHESSI. Following the convention, here 
we refer to the time period before the peak of the SXR 1 -- 8~\AA\ light curve as the {\em rise phase } or the {\em impulsive phase} of a flare, followed by the {\em gradual phase}, or the {\em decay phase}.

Most of these flares exhibit the well-known Neupert effect, namely the flare HXR light curve is temporally
correlated with the time derivative of the 1 -- 8~\AA\ SXR light curve $\dot{\mathcal{F}}_{{\rm sxr}}$ during the rise of the SXR emission. 
To examine the degree to which the Neupert effect applies, we conduct a time-lagged cross-correlation between $\dot{\mathcal{F}}_{{\rm sxr}}$
and the HXR light curves at 12 - 25~keV and 25 - 50~keV, respectively, and the derived maximum cross-correlation coefficients
and time lags are given in Table~\ref{flareinfo}. In a few flares, the HXR emission in 12 - 25~keV lags
$\dot{\mathcal{F}}_{{\rm sxr}}$ by within a minute, likely due to the mixture of thermal emission in this channel \citep[e.g.,][]{Veronig2005,
McAteer2013}. In comparison, the HXR emssion in 25 - 50~keV (not shown in the figure) does not lag $\dot{\mathcal{F}}_{{\rm sxr}}$. 
Since most of these flares do not exhibit significant HXR emissions beyond 25~keV, here we do not conduct a comprehensive energy-dependent analysis
\citep[e.g.][]{McAteer2013}; instead, this study focuses on flare UV light curves in the AIA 1600\AA\ passband.

Readers are reminded that, throughout the following text, the flare UV light curve, $\mathcal{F}_{{\rm uv}}$, specifically refers to 
emission in the AIA 1600~\AA\ passband. The flare emission in this passband is dominated by C~{\sc iv}, Si~{\sc ii}, C~{\sc i}, and 
He~{\sc ii} lines formed in the transition region and the upper chromosphere in the temperature range $4.2 < {\rm log}T < 5.1$
\citep{Simoes2019}. Using high-resolution spectral observations by the Skylab during the decay
phase of a flare, \citet{Simoes2019} found that the most notable line, the C~{\sc iv} line (100,000 K) in this
passband, contributes to 26\% of the AIA 1600~\AA\ flare emission. 
Figure~\ref{fig:lightcurve} shows that $\mathcal{F}_{{\rm uv}}$ matches very well $\dot{\mathcal{F}}_{{\rm sxr}}$ during 
the rise phase, and the coefficients of the cross-correlation and time lags between the two are similar to those between 
$\dot{\mathcal{F}}_{{\rm sxr}}$ and the HXR 12 - 25~keV emission, suggesting a close relation between the HXR emission
and the transition-region and upper-chromosphere line emission \citep[e.g.,][]{Cheng1984}, such as the emission in the AIA 1600\AA\
passband analyzed in this study. On the other hand, it is noted that the flare UV emission 
at this passband proceeds for a longer time than both the $\dot{\mathcal{F}}_{{\rm sxr}}$ and HXR light curves. 

The flare emission in the AIA 1600\AA\ passband is produced by heating of the transition region or upper chromosphere with 
reconnection released energy carried along newly formed flare loops into the lower atmosphere at their feet.
Figure~\ref{fig:2event} shows, as examples, two flares SOL2014-04-18 (event \# 7) and 
SOL2013-08-12 (event \# 4), respectively. The left panels show the evolution of flare ribbons 
in the UV 1600\AA\ passband mapped on a line-of-sight magnetogram obtained from the 
{\em Helioseismic and Magnetic Imager} \citep[HMI; ][]{Schou2012}. The color code indicates the earliest
time a pixel is brightened, or its activation time, defined as the time when its brightness reaches 
4 times the pre-flare quiescent background \citep{Qiu2010}.
The right panels show the UV 1600\AA\ light curves from a few brightened pixels during the flare. From these figures, it is evident 
that, after the impulsive phase of a flare, reconnection continues to form flare loops and releases energy in them,
and the continuous reconnection into the decay phase contributes to the prolonged total UV emission.
These observations suggest that spatially resolved flare light curves of UV or optical emission in the lower atmosphere
provide a comprehensive temporal coverage and spatial mapping of reconnection energy release events in a flare. Therefore, 
in this study, we use the flare UV 1600\AA\ emission 
as the proxy for flare heating regardless of the heating mechanism. We examine 
the Neupert effect that relates spatially resolved UV light curves with the total
SXR light curve, and estimate heating energies in flare loops assumed to be anchored
at the UV-brightened pixels.

For this purpose, we obtain spatially resolved UV 1600\AA\ light curves in flaring
pixels whose brightness is increased to at least 4 times the quiescent background and stays
bright for at least 4 minutes. The first criterion is used to distinguish flaring pixels
from plages, whose brightness distribution peaks at 3.5 times the quiescent background.
The second criterion helps to pick out pixels at the feet of closed loops, different from the feet of open field
lines, or ejecta, which are brightened only briefly. For each of the flares in Table~\ref{flareinfo}, a few thousand
flaring pixels are identified. We assume that, anchored to each UV bright pixel
is a flaring half-loop, and the UV brightness at the pixel is somewhat scaled to the heating flux
in the half-loop. In the foregoing text, each of these half-loops is called a {\em loop event} or a {\em heating event}. 
We then use two methods, an empirical formula of Neupert effect and a zero-dimensional hydrodynamic code, to model 
these heating events and reproduce the synthetic SXR light curve $\mathcal{F}_{{\rm sxr}}$ comparable with GOES observations.

We specify the time range for the analysis of the UV 1600\AA\ and SXR light curves. The start time
$t_s$ of a flare is defined as when $\mathcal{F}_{{\rm sxr}}$ rises to $e^{-4}$ of its peak emission.
The end time of the flare $t_e$ is defined by the $\mathcal{F}_{{\rm uv}}$, instead, as when $\mathcal{F}_{{\rm uv}}$ 
decays to $e^{-2}$ of its maximum. 
The duration of the flare is $\tau_d = t_e - t_s$, and is reported in Table~\ref{flareinfo}.

\section{NEUPERT EFFECT: AN EMPIRICAL MODEL}
The Neupert effect refers to the observation that the time-integrated HXR or microwave light curve matches the SXR light curve
from its rise to peak. The SXR emission then decays because of the reduced emissivity in the passband 
due to decreased temperature (cooling) and/or density, which is not addressed by the Neupert effect in its original form. 
Furthermore, during a flare, numerous flare loops are formed and heated, and then cool, at different times.
The total SXR emission at any given time is the sum of the emissions from these loops, each 
at its own distinct evolution stage; earlier formed flare loops may be cooling during the rise of $\mathcal{F}_{\rm sxr}$, 
whereas new heating events may still take place when $\mathcal{F}_{\rm sxr}$ appears to decay. 

To model the Neupert effect in its complete form, we take into consideration the discrete nature of flare heating as well as cooling
in individual flare loops, and compare the sum of the flare emission from multiple loops with the observed total SXR 
emission. We assume that each newly brightened UV pixel is the foot of a newly formed flare half-loop, and the UV light curve
of the pixel is simply scaled to the heating rate in the loop event. We then convolve the UV light curve of each loop event with a 
kernel function $\mathcal{K}$ that represents the decay of the flare emission in the loop. The modeled total SXR 
emission is therefore given by
\begin{equation}
\mathcal{F}_{{\rm sxr}}(t) = c_0 \sum_{i = 1}^N \int_0^t \mathcal{F}_{{\rm uv},i}(t') \mathcal{K}_i(t, t') dt', 
\end{equation}
where subscript $i$ indicates the contribution from the $i$th loop event, assumed to be anchored to the $i$th UV brightened
pixel. $c_0$ is a scaling constant relating SXR and UV emissions. We have experimented with several forms of the kernel 
function, and found that the function of a half-Gaussian provides the best model:
\begin{equation}
\mathcal{K}_i(t, t') = {\rm exp}\left[\frac{-(t - t')^2}{2\tau_i^2}\right]	 (t > t'),
\end{equation}
where $\tau_i$ is the decay timescale of the emission of the $i$th loop event. When 
$\tau_i \rightarrow \infty$, Equation 1 gives the traditional description of the Neupert effect, that $\mathcal{F}_{{\rm sxr}}$ is the
time integral of $\mathcal{F}_{{\rm uv}}$ without taking into account cooling. 

An automated routine is run to search for the optimal decay timescale $\tau_i$ so that the model light curve 
$\mathcal{F}_{{\rm sxr}}$ matches the observed light curve. Our experiments suggest 
that Equation 1 with a same constant $\tau_i$ for all loop events cannot reproduce the observed $\mathcal{F}_{{\rm sxr}}$ from
rise to decay. We then allow the decay time $\tau_i$ to be time-dependent, considering that, as the flare evolves, 
reconnection takes place at higher altitudes producing longer loops, which take a longer time to cool. For a 
given flare, we use the following trial function to determine $\tau_i$ 
\begin{equation}
\tau_i = \tau_0 {\rm exp}\left[\frac{t_i - t_s}{f\tau_d}\right].
\end{equation}
Here $t_i$ is the peak time of $\mathcal{F}_{{\rm uv}, i}$ for the $i$th loop event, $t_s$ and $t_e$ are the 
start and end times of the flare previously defined, and $\tau_d \equiv t_e - t_s$ is the duration of the flare. 
For each loop event, $\tau_i$ is constant. For each flare, $\tau_0$ and $f$ are constant, 
which give the decay time at the start of the flare and the growth rate of the decay time as the flare evolves. 
For each flare, the automated routine searches for the optimal set of $\tau_0$,  $f$, and $c_0$ that produce the 
best overall correlation and smallest deviations between the model and observed $\mathcal{F}_{{\rm sxr}}$ 
during the time period from $t_s$ to $t_e$. 

\begin{figure}
\epsscale{1.2}
\plotone{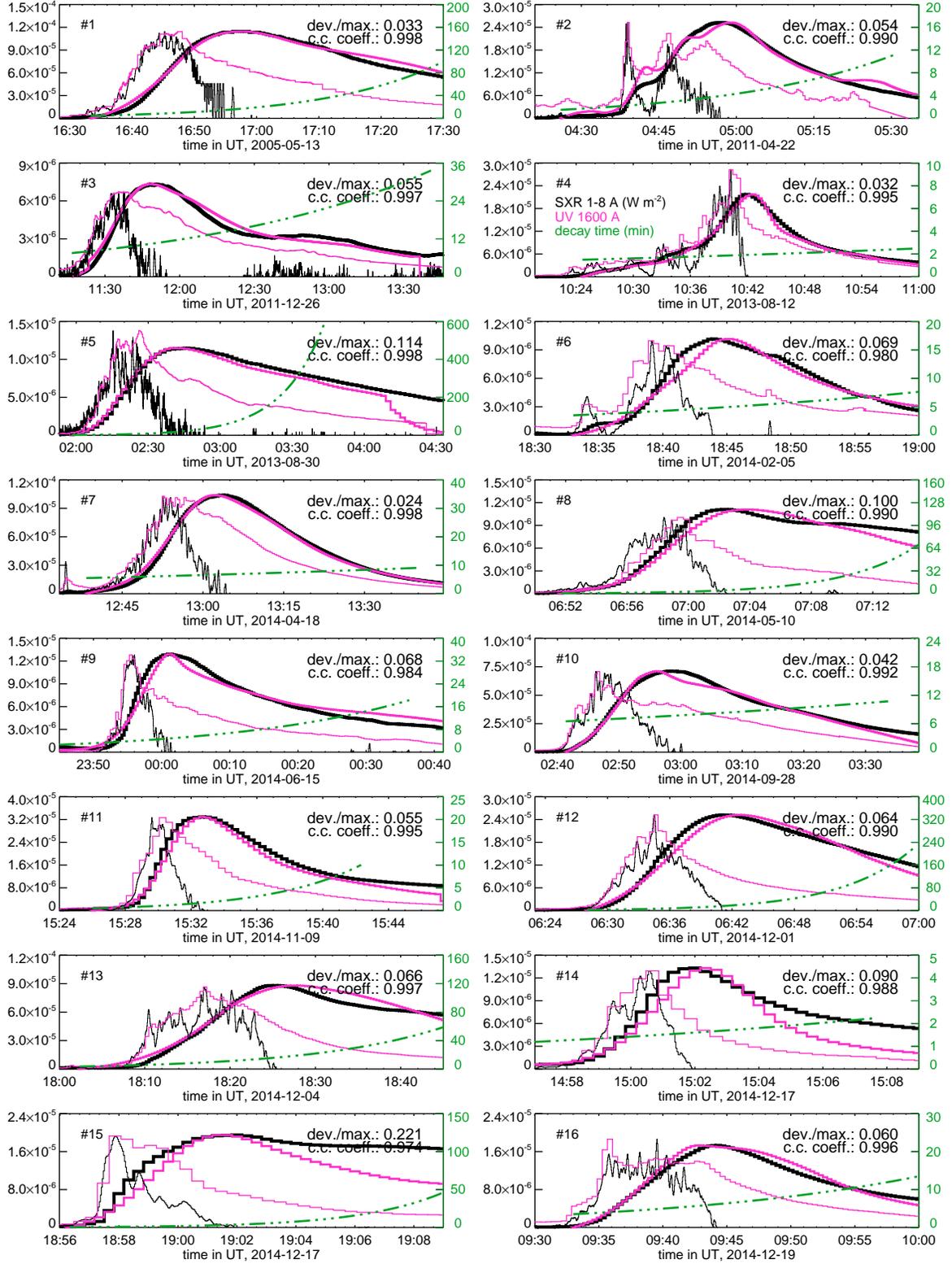}
\caption{Comparison of the observed SXR ``true" flux light curve in 1 -- 8~\AA\ (thick black) with the SXR light curve generated
by the empirical model of the Neupert effect (thick pink). Thin curves show the time derivative of 
the observed SXR light curve (black) and the observed total UV light curve in AIA 1600~\AA\ (pink), both arbitrarily 
scaled. The green curve shows the time dependent decay timescale $\tau_i$ in minutes (see text). 
Also marked are the variance (normalized to the observed peak SXR emission) and the 
coefficient of the cross-correlation between the model and observed SXR light curves.} \label{fig:neupert}
\end{figure}

Figure~\ref{fig:neupert} shows the comparison of the model (thick solid pink) and 
observed (thick solid black) $\mathcal{F}_{{\rm sxr}}$ for the 16 flares analyzed in this 
paper. Also shown in thin solid lines are the total light curve in the AIA 1600\AA\ passband
$\mathcal{F}_{{\rm uv}}$ (pink) and the time derivative of $\mathcal{F}_{{\rm sxr}}$ (black). 
Seen from the figures, the majority of the flares are very well modeled by Equation 1, and
the mean difference between the model and observation normalized to the peak of $\mathcal{F}_{{\rm sxr}}$ is within 10\%. 
Events \#14 and \#15 are the least successful, suggesting that the flare evolution in these two events 
may deviate from the general description by Equation 1, particularly in the decay phase. 
The overall success of this simple model in the majority of the flares suggests that hydrodynamic 
evolution of flare loops, which contribute to the GOES 1- 8\AA\ SXR emission, may be governed by some general rules 
\citep{Warren2004}.

Also shown in Figure~\ref{fig:neupert} is the variation of $\tau_i$ (green) as the flare evolves. Except 
for event \# 4, a growing decay timescale is required to reproduce both the rise and decay of 
the total SXR emission. Qualitatively this is consistent with the general observation that,
as flare evolves, reconnection takes place at higher altitudes, forming longer loops, which cool more slowly.
Observations show the growing separation of the two ribbons (e.g. Figure~\ref{fig:2event}a), an evidence
for growing loops. However, in a few flares (e.g. \# 5), during the decay of the SXR emission,
$\tau_i$ becomes much longer than expected cooling timescales based on observed flare lengthscales and typical
thermodynamic properties of flare loops. Therefore, the empirical {\em decay} timescale found here 
to match the observation is not necessarily the same as the {\em cooling} timescale. 

We also note that the empirical model (Equation 1) has also been applied to HXR light curves (in which case $N = 1$), or
the impulsive component of $\mathcal{F}_{uv, i}$ with its slow-decay component truncated, but cannot 
produce a good agreement with observed $\mathcal{F}_{sxr}$. These experiments indicate that continuous heating in the gradual
phase seems essential in individual loop events and throughout the flare evolution \citep{Qiu2016, Zhu2018}. 
The empirical model supports the scenario requiring the gradual phase heating in individual loop events, but the model itself
is not physical and cannot return the heating rates. To find the amount of energy used in heating the flare corona, we then 
employ the UFC method to model evolution of flare loops.

\section{NEUPERT EFFECT: THE UFC METHOD}
\begin{figure}
\epsscale{1.2}
\plotone{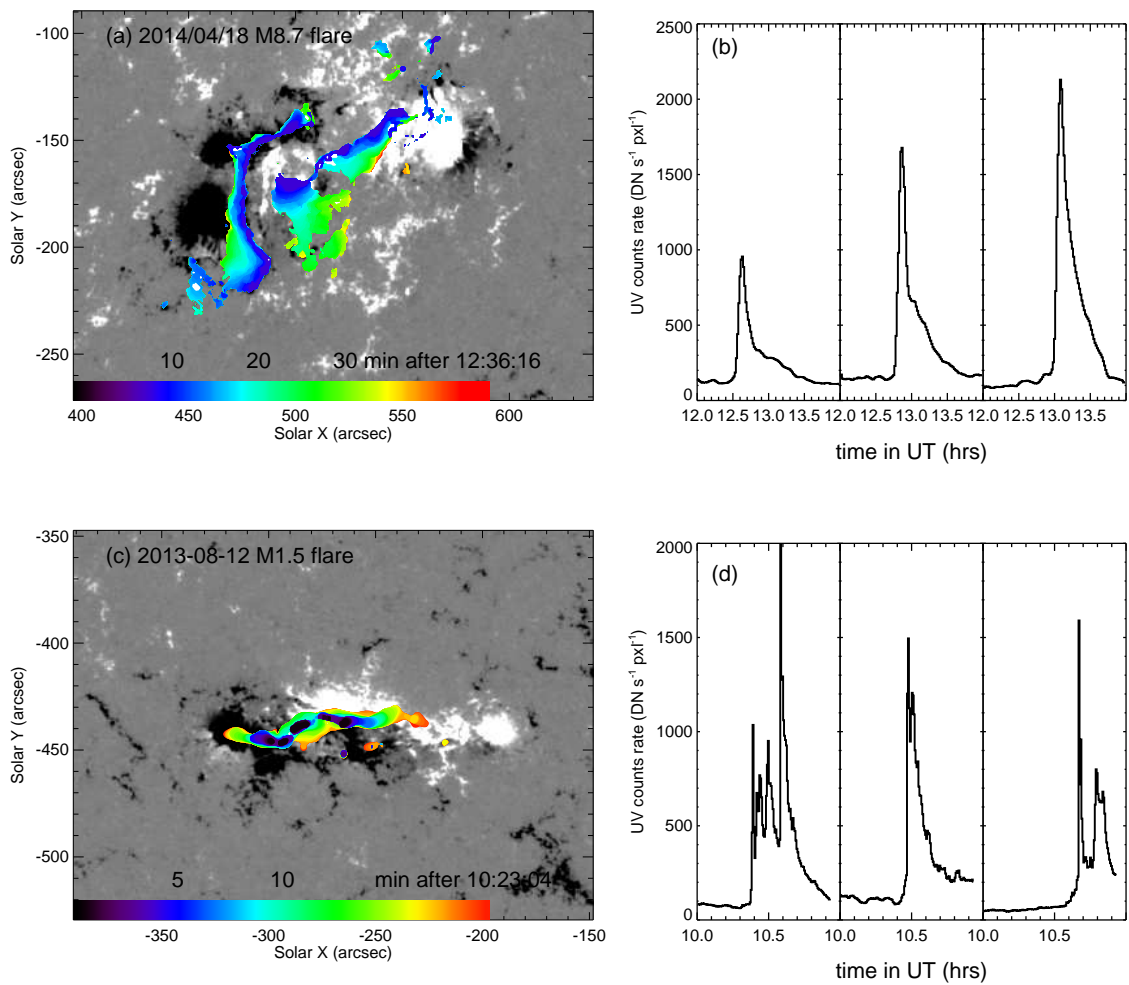}
\caption{Left: evolution of flare ribbon brightening in UV 1600\AA\ passband superimposed on a line-of-sight
magnetogram, obtained by HMI, for the flare \#7 SOL2014-04-18 (a), and the flare \#4 SOL2013-08-12 (c). 
For display, the magnetogram is saturated at $\pm$300 G. The color code indicates the time of the 
start of the flare brightening defined as when the brightness is 4 times the brightness of the pre-flare
quiescent background. Right: UV 1600~\AA\ light curves in a few brightened pixels, showing that flare energy release takes 
place in different places (loops) at different times and proceeds into the decay phase of the flare SXR emission.} \label{fig:2event}
\end{figure}

The encouraging result from the modified empirical model of the Neupert effect indicates that spatially resolved UV emission may be used as 
a proxy for heating rates in flare loops. \citet{Qiu2012, Liu2013} have implemented this idea, and developed the UFC
method to model flare heating. The method infers heating rates in loop events from the UV
lightcurves at the foot-points and models plasma evolution in these loop events 
with a zero-dimensional hydrodynamic code, the Enthalpy-based Thermal Evolution of Loops
model \citep[EBTEL; ][]{Klimchuk2008, Cargill2012b}. The UFC method has been applied to analyze and model seveval flares 
with varying degrees of success \citep{Qiu2012, Liu2013, Zeng2014, Qiu2016, Zhu2018}. The latest effort by \citet{Qiu2016} and \citet{Zhu2018} has suggested
that, even in one loop event, heating takes place in two phases, an intense impulsive heating phase
lasting for a few minutes followed by a gradual heating phase lasting for up to a few tens of minutes yet at a much lower rate. These two phases of heating are reflected
in the UV light curve of a single pixel (see Figure~\ref{fig:2event}b), usually exbihiting a sharp impulsive rise followed
by a long decay. Therefore, in the latest experiment, the UV light curve has been used to infer the heating rate in both
the impulsive and gradual phases of heating, with which, \citet{Zhu2018} have successfully modeled
a two-ribbon flare with the model synthetic emissions in agreement with the observed emissions in 15 passbands by
GOES, AIA, the {\it Extreme-ultraviolet Variability Experiment} \citep[EVE; ][]{Woods2012}, and the {\it X-ray Telescope} \citep[XRT;][]{Golub2007}.

In this paper, we use the UFC method to model the 16 flares with a specific focus on understanding the relationship between UV light curves
in AIA 1600\AA\ passband and GOES SXR lightcurves. The details of the method are given in \citet{Qiu2012, Liu2013}, with the most recent update 
by \citet{Zhu2018}, which takes into account the two-phase heating as well as an empirical treatment
of thermal conduction suppression \citep{Jiang2006}. In this study, we apply this updated model with the empirical term of turbulent 
suppression of thermal conduction, which gives rise to higher plasma temperature at the peak of the flare heating. 
For simplicity, we do not aim at the full-scale comparion of the model 
results with multi-passband observations as done before, but focus on the GOES SXR light curves at 1 -- 8~\AA\ and 0.5 -- 4~\AA.
In addition, we also constrain the cooling rates by comparing the model results with the light curves from the AIA 211\AA\ passband,
which captures flare emission at 2~MK as plasma cools down.
For each flare, we use a scaling constant $\lambda$ to convert observed data counts of the UV 1600\AA\ light curve 
of a brightened pixel to energy flux in the corresponding loop event: $\mathcal{Q}_i(t) L_i = \lambda \mathcal{F}_{uv,i}(t)$, where $\mathcal{F}_{uv, i}(t)$
is the UV 1600\AA\ light curve (in units of DN s$^{-1}$ pxl$^{-1}$), $\mathcal{Q}_i(t)$ is the volumetric heating rate 
(in units of erg cm$^{-3}$ s$^{-1}$), and $L_{i}$ is the length of the half-loop. 
The length of a given half-loop is $L_i = L_0 + v (t_i - t_s)$, $t_i$ being the time
when $\mathcal{F}_{uv,i}(t)$ peaks, and $L_0$ and the growth rate $v$ are estimated from the time-dependent separation of
newly brightened two flare ribbons in the positive and negative magnetic fields, assuming that the half-loop
is a quarter of a circle whose diameter is the mean distance between two flare ribbons. 
With these heating rates as input, and another free parameter $\eta$ that describes the radiative loss from the 
transition region as scaled to the mean pressure in the flare loop \citep{Qiu2013}, the model computes the mean temperature 
and density of thousands of loop events that evolve with time, and the resultant time-dependent differential emission measure 
is convolved with the emissivity and instrument response functions.\footnote{The GOES response function is derived with the SSWIDL
code goes\_fluxes.pro, and the response functions for the AIA EUV passbands are derived with aia\_get\_response.pro. These response
functions are provided by the instrument teams using the latest calibration, as of 2020 October, with CHIANTI 9.0.1 atomic database and coronal abundance.
The AIA response functions are also calibrated with EVE.}
For a given flare, $\lambda$ and $\eta$ are constant for all the identified loop events; for different flares, $\lambda$ and $\eta$ may 
be different. We model each flare with varying $\lambda$ and $\eta$ and find the optimal values that give the best comparison between the observed and synthetic
GOES SXR fluxes at two channels and EUV flux at the AIA 211~\AA\ passband.

\begin{figure}
\epsscale{1.2}
\plotone{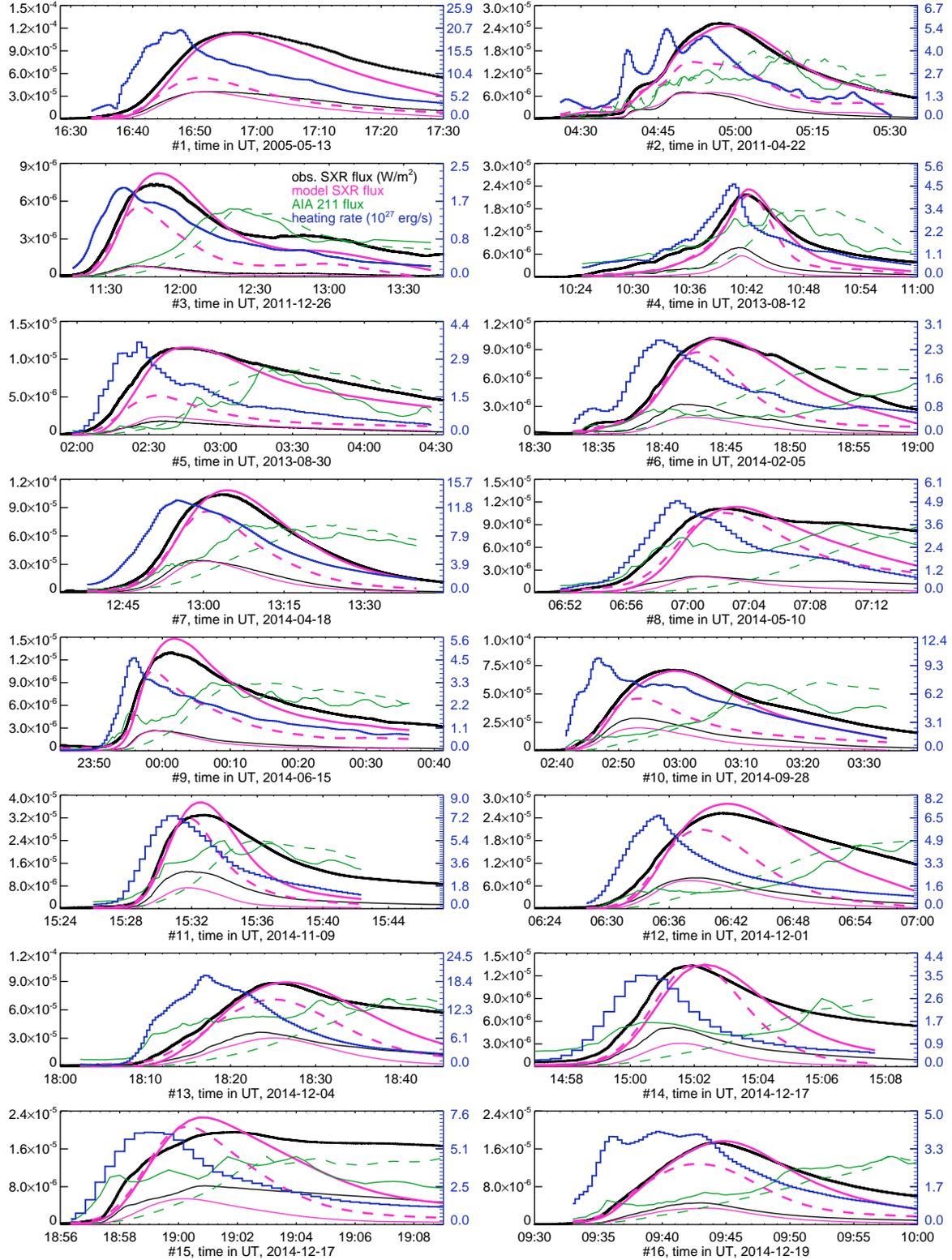}
\caption{Comparison of the GOES observed SXR light curves in 1 -- 8~\AA\ (thick black), 0.5 -- 4~\AA\ (thin black), and
the AIA observed EUV flux at 211~\AA\ passband (solid green), with the synthetic SXRs (thick and thin solid pink)
and EUV (dashed green) light curves by the UFC method that includes gradual heating. For comparison, the SXR 1 -- 8~\AA\ light curve
by the UFC method using only impulsive heating is shown in thick dashed pink. Also plotted in each panel is the total heating rate (blue) derived
from the UFC method. The AIA 211~\AA\ light curves are arbitrarily scaled. For clarity of the display, uncertainties in the synthetic SXR and EUV light curves and in the heating rates 
are not plotted, but they are described in the text (Section 4).}\label{fig:ebtel}
\end{figure}

Figure~\ref{fig:ebtel} shows the comparison of the observed and model synthetic SXR and EUV fluxes for the 16 flares. In each panel, 
the synthetic SXR light curves in 1 -- 8~\AA\ (thick solid pink) and 0.5 -- 4~\AA\ (thin solid pink), 
and EUV 211~\AA\ light curve (dashed green) are average from two model runs conducted with different $\lambda$ and $\eta$ values 
that produce the optimal comparison with observed SXRs (solid black) and EUV 211~\AA\ flux (solid green). The total heating
rate (blue) is also the average of the two runs. 
For clarity of the display, the synthetic and observed GOES SXR flux in 0.5 -- 4~\AA\ is multiplied by a factor of two, and uncertainties, which are small fractions of the
mean fluxes, are not plotted in the figure.
Seen in the figure, in the majority of the flares, the synthetic SXR and EUV fluxes are in reasonable agreement with the observed fluxes.

Note that the zero-dimensional model is not capable of accurately calculating plasma properties out of equilibrium during the very
dynamic heating phase in the first few minutes; therefore, the model cannot produce sufficient SXR 0.5 -- 4~\AA\ emission at very high temperatures,
which is likely the case in a few flares, like event \# 11. Nevertheless, the total SXR 1 -- 8~\AA\ and EUV emissions summed over all loops
during the flare timescale are mostly produced at lower temperatures, and they much depend on the total energy deposit 
in the loops and are less subject to the details of heating and plasma evolution in non-equilibrium in the short impulsive heating phase 
\cite[see discussions by ][]{Winebarger2004}. Therefore, the overall agreeable comparison between the synethic and observed total fluxes 
suggest that the heating rates inferred from the flare foot-point UV 1600\AA\ emissions are reasonable first-order estimates. It is 
noted, though, that in the decay phase of a number of flares, the model does not produce sufficient $\mathcal{F}_{sxr}$ emission as observed. 
This will be further discussed in the next section, in conjuction with the result of the empirical Neupert model. 

We remind that the profile of the heating rate for each loop event used in the model resembles the time profile 
of the UV light curve at the foot, which generally consists of an impulsive component followed by a gradual component 
(see Figure~\ref{fig:2event}). As a comparison, the thick dashed pink curves in Figure~\ref{fig:ebtel} show the synthetic 
$\mathcal{F}_{sxr}$ in 1 -- 8~\AA\ with the impulsive heating model. For the impulsive model, the heating rate of a 
loop event is derived by fitting the rise of the UV light curve to a half gaussian, and the impulsive heating rate is a full gaussian \citep{Qiu2012, Liu2013}. 
All other properties, such as the lengths of the loop events, are the same in the impulsive heating model and two-phase heating model.
The figure shows that, in the majority of the flares, the two-phase heating model produces synthetic SXR emissions in much better agreement 
with the observed SXR emission than the model only using impulsive heating rates. 
The necessity of two-phase heating requires a greater amount of flare heating
energy than the impulsive heating. 
In different flares, the fraction of impulsive heating energy out of the 
total varies from 40\% to 85\%; on average, the amount of heating energy in the impulsive components 
takes up about two thirds of the total heating energy, and the remaining one third of heating energy 
is distributed in the gradual components of the heating events. 

\section{ENERGETICS OF FLARE HEATING}

\subsection{Estimate of Flare Heating Energy}
The UFC method allows us to estimate the total energy deposit in the flare corona. However, in a number
of flares, the model still does not produce sufficient SXR emission in the decay phase; therefore, the total heating energy derived
directly from the UFC method is likely the lower-limit of the corona heating energy. On the other hand, 
$\mathcal{F}_{sxr}$ produced by the empirical model compares better with the observation in the decay phase;
yet the empirical model only relates the time evolution of flare SXR and UV 1600\AA\ emissions, and cannot return the heating
rates. To help improve estimates of heating energies, we may use the results from the empirical 
model to calibrate heating energies derived from the UFC method. 

\begin{figure}
\epsscale{1.2}
\plotone{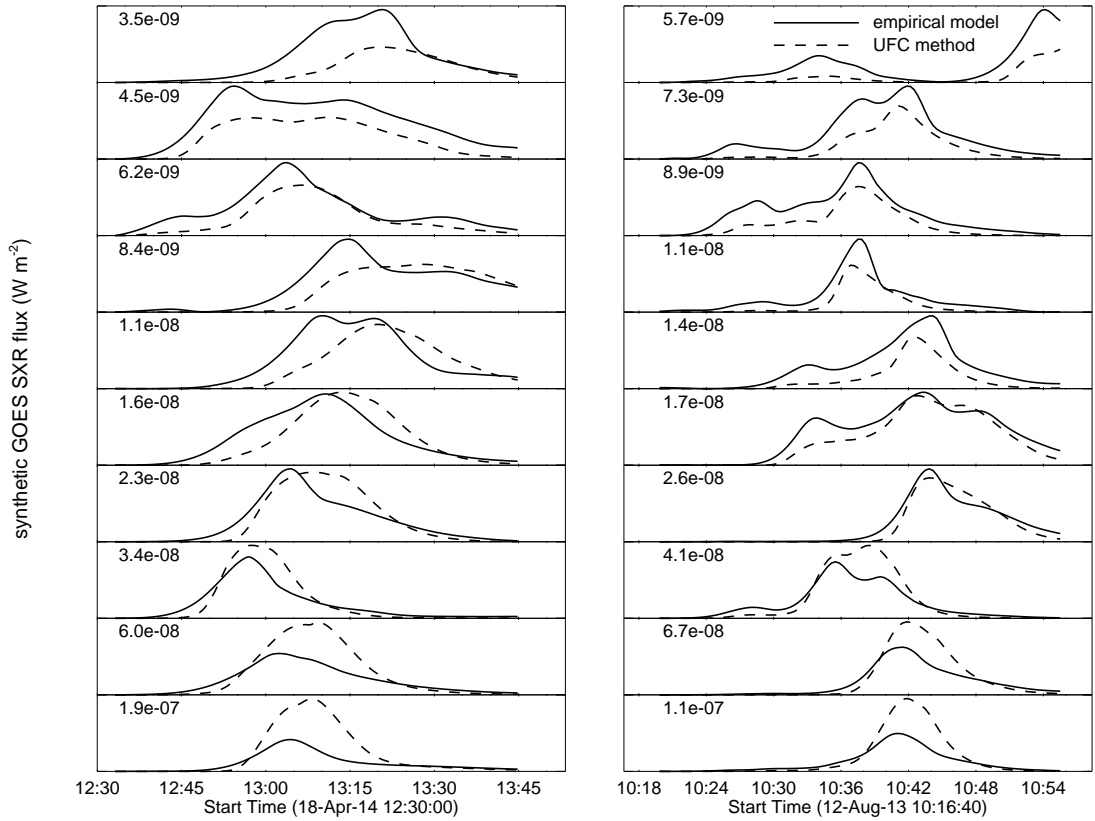}
\caption{Left: synthetic SXR light curves in 1 -- 8 ~\AA\ with the empirical model (solid) and UFC method (dashed), respectively, 
in 10 randomly sampled loop events for the flare SOL2014-04-18. Right: same as the left but for the flare SOL2013-08-12.
Marked in each panel is the peak flux of the SXR light curve by the UFC method. 
}\label{fig:single_gflx}
\end{figure}

To understand the difference in the total SXR flux produced by the two models, we compare the
synthetic SXR flux in individual loop events. Figure~\ref{fig:single_gflx} shows the synthetic 
SXR light curves in ten randomly sampled loop events generated by the empircal model (solid) 
and the UFC method (dashed), respectively, for the two flares displayed in Figure~\ref{fig:2event}. 
It is seen that the SXR flux generated by the two models have very similar time profiles, yet for 
weak events, the magnitude of the SXR flux by the UFC method
is lower than that by the empirical model. Such comparison may explain the insufficient SXR emission by the UFC method
during the decay of the flare, when flare heating and the SXR flux in individual loop events become
smaller. Since the empirical model is able to produce the total SXR flux which 
compares better with the observation, we will assume that the SXR emission in each loop event generated by the 
empirical model represents the ground truth, and uses it to make new estimates of heating energies in flare loops.

\begin{figure}
\epsscale{1.2}
\plotone{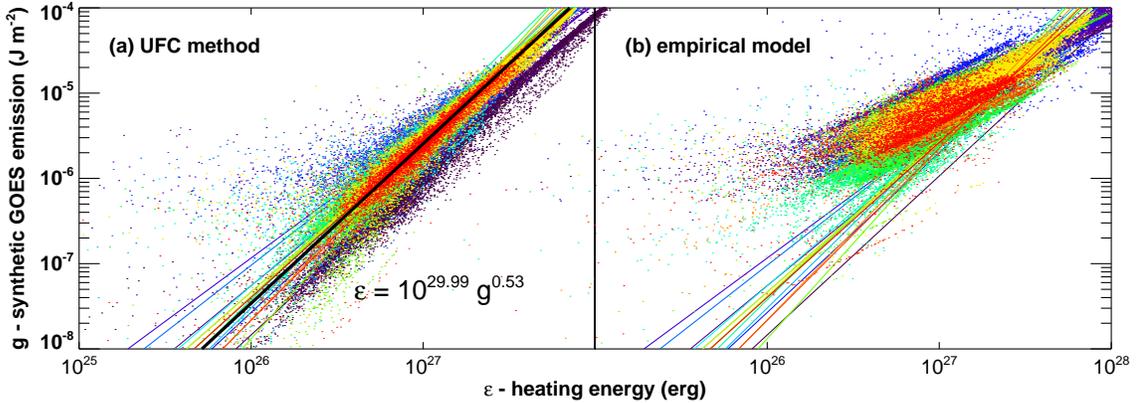}
\caption{Scatter plot of the time integrated SXR flux in 1 -- 8~\AA\ $\mathcal{G}$ generated by the UFC method (a) 
or the empirical Neupert model (b) against the total heating energy $\mathcal{E}_{ufc}$ in individual loops. 
Each color shows a few thousand loop events for a given flare, and the solid line of the same color illustrates 
the $\mathcal{E}_{ufc} - \mathcal{G}_{ufc}$ fit to a power law for the same flare. The black solid line shows 
the $\mathcal{E}_{ufc} - \mathcal{G}_{ufc}$ fit to a power law for all loop events in all 16 flares. Note that
the solid color lines in (b) are the same as in (a), for comparison of the synthetic SXR emissions 
generated by the two models.}\label{fig:single_loop}
\end{figure}

For this purpose, we first establish the relation between the heating energy and the synthetic GOES SXR emission
by the UFC-EBTEL model. The left panel of Figure~\ref{fig:single_loop} shows a scatter plot of the time 
integrated heating energy in the loops, 
denoted as $\mathcal{E}_{ufc}$ (in units of erg), versus the time integrated synthetic GOES SXR flux generated by the EBTEL model,
denoted as $\mathcal{G}_{ufc}$ (in units of J m$^{-2}$). The $\mathcal{E} - \mathcal{G}$ scatter plot is quite tight for each flare, and 
can be described by a power-law $\mathcal{E} \approx 10^{\beta} \mathcal{G}^{\alpha}$. For the flares modeled in this event, 
$\alpha$ ranges between 0.45 and 0.67, and $\beta$ ranges from 29.52 to 30.56. In fact, the $\mathcal{E} - \mathcal{G}$ relation for all 
loop events in all 16 flares can be fitted to one power-law, as shown in the figure (solid black line),
yielding $\langle\alpha\rangle = 0.535\pm0.001$ and $\langle\beta\rangle = 29.990\pm0.004$.
This scaling law allows us, without running the hydrodynamic model, to estimate the total SXR emission in
a loop event given the amount of the heating energy, and vice versa. 

In comparison, the right panel of Figure~\ref{fig:single_loop} shows the time integrated synthetic SXR emission generated by the empirical model
$\mathcal{G}_{emp}$ -- for a better comparison, we exclude event \#14 and \#15 that 
are not well modeled with the empirical formula. As expected, $\mathcal{G}_{ufc}$ becomes increasingly under-estimated 
for smaller $\mathcal{G}_{emp}$. Based on these analyses, we make a new estimate of flare heating energy, denoted 
as $\mathcal{E}_{emp}$, using $\mathcal{G}_{emp}$ as the ground truth for the SXR emission by each loop event to
replace $\mathcal{G}_{ufc}$ in the $\mathcal{E} - \mathcal{G}$ scaling, namely, $\mathcal{E}_{emp} \approx 10^{\beta} \mathcal{G}_{emp}^{\alpha}$.
The estimate can be made using $\alpha$ and $\beta$ derived for each flare, or $\langle\alpha\rangle$ and $\langle\beta\rangle$
derived for all flares, and the difference in the estimate is not found significant. We take the average
$\mathcal{E}_{emp}$ from these two estimates as a plausible upper-limit of the heating energy in each loop event, whereas
the heating energy $\mathcal{E}_{ufc}$ derived from the original UFC method is taken as the lower limit. 

\begin{figure}
\epsscale{1.2}
\plotone{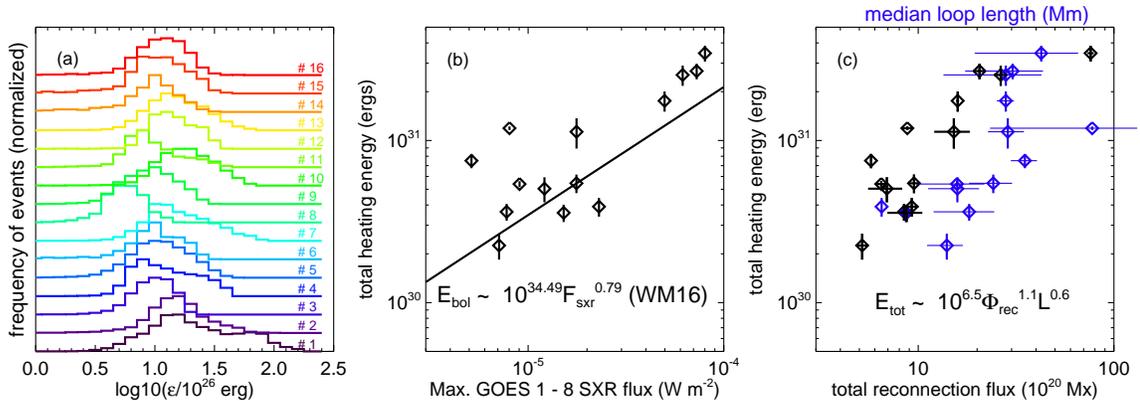}
\caption{(a): histograms of the heating energies in the loop events for each of the 16 flares analyzed in the paper. Here the heating energy in
each loop event is the average of $\mathcal{E}_{ufc}$ and $\mathcal{E}_{emp}$. (b) Scatter plot of the total heating energy
against the magnitude of the flare (based on the ``scaled" flux). Vertical bars indicate the range of the total heating energy, with
$\sum\mathcal{E}_{ufc}$ being the lower limit and $\sum\mathcal{E}_{emp}$ being the upper limit. The solid guide line shows the power-law scaling
of the observed bolometric radiation energy to the flare magnitude given by \citet{Warmuth2016}.  
(c) The total heating energy against the reconnection flux $\Phi_{rec}$ (black; see text) and median length $L$ of the flare
loop events (blue). Vertical bars indicate the ranges of the flare heating energy as in (b); horizontal bars indicate the uncertainties
of the $\Phi_{rec}$ measurements (black) or the standard deviations of the estimated lengths (blue) of the loop events that
are subsequently formed during the flare evolution from rise to decay.}\label{fig:energy}

\end{figure}

Figure~\ref{fig:energy}a shows the distribution of heating energies $\mathcal{E}$, the mean 
of $\mathcal{E}_{emp}$ and $\mathcal{E}_{ufc}$.
The new estimate changes the distribution of heating energies in the loop events, which becomes tighter
toward higher energies, and raises the total flare heating energy by one third on average. 
Panel (b) shows the total energy $\mathcal{E}_{tot} = \sum \mathcal{E}$ (in ergs) that is used to heat the flare corona for each of the 14
flares (i.e., excluding event \#14 and \# 15), plotted against the flare magnitude defined by the peak SXR 
flux in GOES 1 - 8~\AA\ channel\footnote{Here, to be consistent with prior literature, the 
flare magnitude is derived with the ``scaled" GOES SXR flux, but not the ``true" flux. The ``true" flux in this channel, as released
in October 2020, is equivalent to the ``scaled" flux divided by 0.7 (\texttt{https://hesperia.gsfc.nasa.gov/rhessidatacenter/complementary\_data/goes.html}).}.
Each vertical bar indicates the range of the total heating energy, the 
lower limit being the sum of $\mathcal{E}_{ufc}$ and the upper limit the sum of $\mathcal{E}_{emp}$,
and the symbols indicate $\mathcal{E}_{tot}$, the mean of $\sum\mathcal{E}_{ufc}$ and $\sum\mathcal{E}_{emp}$.
Overplotted is the scaling law by \citet[][WM16 scaling law hereafter]{Warmuth2016}
that relates the total (bolometric) radiation energy of flares observed between 1996 and 2007 \citep{Kretzschmar2011, Emslie2012}
to their GOES magnitude: $\mathcal{E}_{bol} \approx 10^{34.49\pm0.44} \mathcal{F}_{sxr}^{0.79\pm0.10}$. The total heating 
energy derived in this study scatters around the WM16 scaling law\footnote{The energy-magnitude scaling in this study is $\mathcal{E}_{tot} \approx 10^{34.33\pm0.81} \mathcal{F}_{sxr}^{0.72\pm0.17}$}, suggesting that 
this study has achieved a close estimate of the total heating energy in flares.
\citet{Warmuth2016} also derived the maximum thermal energy $\mathcal{E}_{th}$ and non-thermal electron energy 
$\mathcal{E}_{nth}$ of 24 flares observed by GOES and RHESSI between 2002 and 2003, which scale with the flare magnitude
as $\mathcal{E}_{th} \approx 10^{33.67\pm0.26} \mathcal{F}_{sxr}^{0.88\pm0.06}$, and $\mathcal{E}_{nth} \approx 10^{35.07\pm0.38}
\mathcal{F}_{sxr}^{1.08\pm0.09}$, respectively. The heating energy estimated here is nearly an order of magnitude larger than the maximum
thermal energy, and is also greater than the non-thermal electron energy, particularly in small flares.
Therefore, flare heating is not entirely due to non-thermal electrons, and the foot-point UV emission signatures
more comprehensively capture heating events during the flare regardless of heating mechanisms.

\subsection{Reconnection and Energetics}
Magnetic reconnection forms flare loops and releases energy that is used to heat flare loops. 
The amount of magnetic flux $\Phi_{rec}$ participating in reconnection is measured by summing up 
the magnetic flux in the pixels (see Figure~\ref{fig:2event}a, c) whose brightness in the 1600~\AA\ passband
is increased to be more than 4 times the quiescent brightness and for at least 4 minutes. 
Flares in this study take place near the disk center, and we integrate the HMI measured longitudinal 
photospheric magnetic flux density $B$ (in units of Gauss, or Mx cm$^{-2}$)
in flaring pixels, without correcting the projection effect and without extrapolating $B$ 
to the upper-chromosphere or transition region, since these two effects
partly cancel each other. Finally, the measurement assumes that each patch of magnetic flux 
anchored at a UV-brightened pixel participates in magnetic reconnection only once to form a flare loop
containing this flux. The uncertainty is estimated from $\Phi_{rec}$ measured in the positive
and negative magnetic fields, which, on average, is about 20\% of $\Phi_{rec}$ \citep[also see][]{Qiu2005, Qiu2007}.

Figure~\ref{fig:lightcurve} shows the reconnection rate $\dot{\Phi}_{rec}$, the time derivative 
of the time-dependent reconnection flux, which varies in the range of $10^{17-19}$ Mx s$^{-1}$
from flare to flare. The figure shows that $\dot{\Phi}_{rec}$ is more impulsive and usually precedes the total heating 
rate $\dot{\mathcal{E}}_{tot}$. In most flares, $\dot{\Phi}_{rec}$ does not diminish to zero after the peak of the SXR emission, 
indicating that reconnection and formation of new flare loops continue into the decay phase,
although at a much smaller reconnection rate and the amount of reconnection flux making only a small fraction
of the total reconnection flux. On the other hand, the analysis of energetics in Section 5.1 suggests
that the total heating energy in the decay phase of the flare is non-negligible, amounting to 27\% on
average. 

These observations imply that the heating energy $\mathcal{E}$ in individual loop events is not a simple linear function
of the magnetic flux in the loop, and loop events in the early phase of the flare have less energy
per unit flux compared with loop events in the later phase of the flare. 
A regression analysis yields a very weak dependence of the heating energy $\mathcal{E}$
on either the magnetic flux or the length of the loop events. 
On the other hand, the integrated total energy of the flare exhibits a much stronger dependence
on the reconnection flux, $\mathcal{E}_{tot} \sim \Phi_{rec}^{1.1\pm0.2} L^{0.6\pm0.1}$, as shown in Figure~\ref{fig:energy}c.\footnote{Note that this
scaling law is derived for the 13 flares all observed by AIA and HMI, excluding event \#1, 14, and 15. Scaling laws
involving magnetic field measurements change significantly when the first event is included. 
With this event included, the energy-flux relation becomes $\mathcal{E}_{tot} \sim \Phi_{rec}^{0.8\pm0.1} L^{0.6\pm0.2}$. 
In addition, the scaling of the flare magnitude and reconnection flux is found to be $\mathcal{F}_{sxr} \sim \Phi_{rec}^{1.6\pm0.2}$
for the 13 flares, similar to that in \citet{Kazachenko2017}, who analyzed more than 3000 flares observed by AIA and HMI and
found $\mathcal{F}_{sxr} \sim \Phi_{rec}^{1.5}$; with the first event included, the magnitude-flux scaling in this study 
becomes $\mathcal{F}_{sxr} \sim \Phi_{rec}^{1.1\pm0.2}$. The first event was observed by TRACE and MDI, so the discrepancy 
might be due to different calibrations of the two generations of the instruments.} 
Here $L$ is the median length of the loop events in units of Mm. The energy dependence on 
$\Phi_{rec}$ is very close to that found by \citet{Reep2019} and \citet{Zhu2018}. \citet{Reep2019}
analyzed a few thousand flares, and the energy in the scaling law refers to the flare thermal energy
at the time of peak temperature, deduced from GOES SXR observations. 
\citet{Zhu2018} analyzed only one event SOL2011-12-16 (\# 3) and the flux-energy patches 
are grouped into a few tens magnetic cells to construct the scaling law; 
\citet{Zhu2018} did not reveal a dependence on the loop length, which does not vary significantly 
during this flare. In a somewhat different context, \citet{Schrijver2004} 
found a similar scaling law $\mathcal{F}_H \sim \langle B\rangle^{1.0\pm0.3}L^{-1.0\pm0.5}$ 
that relates the heating flux $\mathcal{F}_H$ (in units of erg cm$^{-2}$ s$^{-1}$) of active 
regions to the mean magnetic field strength $\langle B\rangle$ at the base (the chromosphere)
and the scale size $L$ of the active region loops. We may re-write their scaling law 
as $\mathcal{E}_{tot} \sim \mathcal{F}_H A \tau \sim \Phi^{1.0} L^{1.0}$,
considering that the magnetic flux is given by $\Phi = \langle B\rangle A$, $A$ being the total cross-sectional area of 
active region loops, and, in equilibrium, the heating timescale is roughly the same as the thermal conduction 
timescale $\tau \sim L^2$. On global scales ranging from active regions to magnetic cells in a given active region,
and within uncertainties, these scaling laws are very similar; in particular, the energy dependence on the 
magnetic field is the same, indicating the similar nature of energy release in these systems \citep{Schrijver2004}. 

\section{CONCLUSIONS AND DISCUSSIONS}
\subsection{Summary}
In this study, we estimate the total energy that is used to heat flare plasmas in the corona using two simplified models, 
an empirical model of the Neupert effect and a zero-dimensional hydrodynamic model (the UFC method). The purpose of the study is 
to derive a first-order estimate of flare energies in a multitude of flare loops.
Although these models are incapable of precisely describing thermodynamic properties during 
the initial impulsive heating phase of a flare loop when non-equilibrium physics governs the loop evolution, 
they are suitable for the thought experiment, as conducted in this paper, on the longstanding perception that 
energy release and flare heating take place in numerous patches over an extended time period. The experiment 
takes advantage of spatially resolved UV emission from the foot-points of flare loops at the transition region or upper chromosphere,
assuming that each UV-brightened pixel represents a single patch of energy release, denoted as a loop event 
or heating event in this study. The experiment extends the traditional concept of the Neupert effect to spatially 
resolved UV light curves. 

We have conducted the experiment on 16 flares ranging from C to M class. The study confirms that a multitude of impulsive 
heating events alone cannot reproduce the observed flare SXR light curve, but the two-phase heating model 
produces the synthetic SXR emission in better agreement with observations. This is consistent with the recent
finding by \citet{Kerr2020}, who have conducted one-dimensional loop simulations with impulsive heating at 
fine scales, and found that the model produced thermodynamic properties decay faster than observed by IRIS. 
Furthermore, comparing the empirical model 
of the UV Neupert effect and the UFC method, the former producing the SXR emission in the decay phase 
in still better agreement with observations than the latter, we have improved the estimate of the flare 
heating energy particularly in the decay phase of the flare; on average, the amount of the heating energy in the 
decay phase of the flare (i.e., after the peak of the total SXR emission in 1 -- 8~\AA) makes 27\% of the 
total heating energy during the flare.

The estimated energies used to heat the flare corona are comparable with the bolometric radiation energy measured in 
flares of similar magnitudes \citep{Warmuth2016}. Therefore, the UV emission signatures at the foot-points
of flare loops well capture heating events during the flare regardless of heating mechanisms. The flare heating energy $\mathcal{E}_{tot}$
is also shown to scale with the total reconnection flux $\Phi_{rec}$ and the median length of the flare half-loops $L$
by $\mathcal{E}_{tot} \sim \Phi_{rec}^{1.1\pm0.2} L^{0.6\pm0.1}$; the dependence of the heating energy on the magnetic field
is similar to scaling laws found in some studies, though with various contexts \citep{Schrijver2004, Zhu2018, Reep2019}, 
but different from some other studies such as by \citet[][ and references therein]{Aschwanden2020a, Aschwanden2020c}. 
On the other hand, we do not find a strong dependence of the heating energy on the magnetic field (flux)
and/or the loop length for individual loop events down to the pixel scale ($\sim$~0.6\arcsec ).

\subsection{Discussions}
Numerous prior studies have examined scaling laws that relate the flare magnitude, namely the peak 
GOES SXR flux in 1 -- 8~\AA, to flare energies of various kinds. Some of these studies also 
take into account the lengthscale of flare loops. Based on the RTV scaling law, 
\citet{Warren2004} found the flux-energy relation to be super-linear $\mathcal{F}_{sxr} \sim \mathcal{E}_{tot}^{1.75}L^{-1}$
(here $\mathcal{F}_{sxr}$ refers to the peak SXR flux in units of W m$^{-2}$), which was confirmed 
with a one-dimensional hydrodynamic model of loop heating by a beam of non-thermal electrons. 
One-dimensional loop simulations by \citet{Reep2013} yielded a similar scaling law $\mathcal{F}_{sxr} 
\sim \mathcal{E}_{tot}^{1.7}$. However, analyzing a few thousand flares using the database by \citet{Kazachenko2017}, 
\citet{Reep2019} found sub-linear scaling laws $\mathcal{F}_{sxr} \sim \mathcal{E}_{th}^{0.85}$,
and $\mathcal{F}_{sxr} \sim \mathcal{E}_{tot}^{0.85}$, the former referring to the 
thermal energy of the flare (at the time of peak temperature) derived from the GOES SXR analysis, and the latter referring to the
flare heating energy deduced from the traditional Neupert effect, i.e., $\mathcal{E}_{tot}$ being 
the non-thermal electron energy. Similarly, \citet{Aschwanden2020b} found $\mathcal{F}_{sxr} \sim \mathcal{E}_{diss}^{0.7}$
where $\mathcal{E}_{diss}$ refers to energy dissipated in flares.
Finally, the scaling laws by \citet{Warmuth2016} would suggest 
$\mathcal{F}_{sxr} \sim \mathcal{E}_{bol}^{1.3}$, $\mathcal{F}_{sxr} \sim \mathcal{E}_{nth}^{0.9}$, 
and $\mathcal{F}_{sxr} \sim \mathcal{E}_{th}^{1.1}$. 

From this study, we find a super-linear flux-energy relation,
$\mathcal{F}_{sxr} \sim \mathcal{E}_{tot}^{1.4\pm0.2}L^{-1.1\pm0.2}$ for 14 flares
(excluding \#14 and \#15 that are not well modeled); again, the flux-energy dependence is closest to 
the WM16 scaling law of the bolometric energy. 
The difference from the other scaling laws by, e.g., \citet{Warren2004, Reep2013, Reep2019, Aschwanden2020b} 
may be due to the fact that flare heating takes place over an extended time period beyond the impulsive phase, 
and is not provided only by non-thermal electrons. 

The modified empirical model of the UV Neupert effect is able to produce SXR light curves in very good agreement
with observations, which is used, in this study, to return an improved estimate of flare energetics, particularly 
in the decay phase. However, we do not fully understand the implication of 
the convolution in the form of a gaussian (Equation 1), with the {\em decay} timescale which becomes very large at times.
Guided by this thought experiment, in the future work, we will investigate the physical reason for the discrepancy
between the two models, and then conduct a full-scale modeling of flare evolution with the improved UFC method
employing multiple-wavelength observations in a larger number of flares (Zhu et al., in preparation).
This study may also serve as a prior experiment for more comprehensive and physics based models, which can
unravel physics of heating mechanisms \citep{Longcope2015, Reep2019b, Kowalski2019,
Graham2020, Kerr2020}, and also help address production of flare UV emissions in the transition region and upper chromosphere
\citep[e.g.,][]{McClymont1986, Milligan2015, Simoes2019}, used in this study as a {\em proxy} for heating.


\acknowledgments The author thanks the referee for constructive comments that help improve the analysis and the clarity
of the manuscript. The auhtor thanks Lilly Bralts-Kelly and Jianxia Cheng for helping prepare the AIA data. 
This work has been supported by the NASA grants NNX14AC06G and 80NSSC19K0269. The work also benefits from
the ISSI/ISSI-BJ team collaboration ``Diagnosing Heating Mechanisms in Solar Flares". 
SDO is a mission of NASA's Living With a Star Program. 

\bibliography{neupert_new}

%
\begin{deluxetable}{llllllllllllll}
\tabletypesize{\scriptsize}
\rotate 
\tablecolumns{10} \tablewidth{0pt} \tablecaption{Properties of Flares and Model Parameters\label{flareinfo}}
    \tablehead{
    \colhead{ } &
    \colhead{start time, magnitude$^{a}$} &
    \colhead{position} &
    \colhead{$\tau_d$} &
    \colhead{$L$} & 
    \colhead{$\Phi_{rec}$} & 
    \colhead{$\mathcal{E}_{tot}$} & 
    \multicolumn3c{cross-correlation coefficient and time lag (sec)$^{f}$} \\
    \colhead{} &
    \colhead{}&
    \colhead{}&
    \colhead{(min)$^b$}&
    \colhead{(Mm)$^c$}&
    \colhead{(10$^{20}$ Mx)$^d$}&
    \colhead{(10$^{30}$ erg)$^e$}&
    \colhead{12-25keV} & 
    \colhead{25-50keV} & 
    \colhead{UV1600}
}
\startdata
1 & 2005-05-13 16:33 M8.0 & NOAA10759 N12E05  & 62  & 43 (23) & 76.2 (5.5) & 34.6 (7.8) & 0.41 (0) & 0.88 (20) & 0.79 (0)\\
2 & 2011-04-22 04:26 M1.8 & NOAA11195 S17E29  & 64  & 29 (6) & 15.2 (6.3) & 11.3 (4.8) & 0.71 (-33) & 0.64 (8) & 0.72 (0) \\
3 & 2011-12-26 11:16 C5.1 & NOAA11384 N13W14  & 160 & 35 (5) & 5.8 (0.1) & 7.5 (1.5) & - & - & 0.65 (-100)\\
4 & 2013-08-12 10:25 M1.5 & NOAA11817 S22E10  & 35  & 9 (0) & 8.7 (3.5) & 3.6 (0.9) & 0.91 (-20) & 0.91 (14) & 0.89 (-40) \\
5 & 2013-08-30 01:58 C8.0 & NOAA11836 N12E28  & 150 & 76 (53) & 8.9 (0.9) & 11.9 (0.6) & 0.76 (0) & 0.53 (208) & 0.74 (-120)\\
6 & 2014-02-05 18:33 C7.1 & NOAA11967 S12W36  & 34  & 14 (3) & 5.2 (0.6) & 2.3 (0.8) & 0.93 (0) & 0.44 (137) & 0.81 (0) \\
7 & 2014-04-18 12:38 M7.2 & NOAA12036 S15W42  & 55  & 31 (13) & 20.6 (3.0) & 26.8 (5.9) & - & - & 0.70 (-100) \\
8 & 2014-05-10 06:52 C7.7 & NOAA12056 N04E17  & 24  & 18 (6) & 8.5 (0.4) & 3.6 (0.9)  & 0.92 (0) & 0.88 (0) & 0.82 (-60)\\
9 & 2014-06-15 23:30 C9.0 & NOAA12087 S18W11  & 66  & 15 (6) & 6.5 (0.9) & 5.4 (0.5) & 0.90 (-41) & 0.85 (127) & 0.93 (-20) \\
10& 2014-09-28 02:41 M5.0 & NOAA12173 S21W24  & 52  & 28 (3) & 15.9 (0.6) & 17.6 (5.0) & 0.72 (-39) & 0.67 (98) & 0.72 (0) \\
11& 2014-11-09 15:26 M2.3 & NOAA12205 N15E05  & 16  & 7 (35) & 9.3 (1.5) & 3.9 (1.0) & - & - & 0.77 (-20) \\
12& 2014-12-01 06:28 M1.8 & NOAA12222 S20E04  & 32  & 25 (6) & 9.5 (1.1) & 5.4 (1.4) & 0.85 (-8) & 0.77 (49) & 0.91 (0) \\
13& 2014-12-04 18:02 M6.2 & NOAA12222 S20W35  & 45  & 28 (15) & 26.4 (4.0) & 25.4 (7.2) & 0.79 (-155) & 0.85 (0) & 0.89 (-20)\\
14& 2014-12-17 14:42 C9.3 & NOAA12242 S19W02  & 25  & 7 (9) & 4.7 (0.1) & 1.1 (0.1) & 0.95 (-24) & 0.98 (10) & 0.96 (-20) \\
15& 2014-12-17 18:56 M1.4 & NOAA12241 S10E17  & 14  & 9 (10) & 8.1 (4.1) & 3.0 (0.8) & 0.37 (-94) & 0.89 (0) & 0.73 (-20) \\
16& 2014-12-19 09:33 M1.2 & NOAA12237 S13W40  & 30  & 16 (5) & 6.9 (2.7) & 5.0 (1.8) & 0.69 (0) & 0.35 (12) & 0.71 (0)\\ 
\enddata
\tablenotetext{a}{Flare magnitude is based on the ``scaled" GOES SXR flux in 1 -- 8~\AA, but not the ``true" flux released
in October, 2020. Determination of the start time $t_s$ is described in the text (Section 2).}
\tablenotetext{b}{The duration of the flare, $\tau_d = t_e - t_s$, where $t_s$ and $t_e$ are start
and end times defined in the text (Section 2).}
\tablenotetext{c}{The median length of flare half-loops. Also shown in the parenthesis is the standard deviation
of the length of the loop events, which grows as the flare evolves (see text in Section 4).}
\tablenotetext{d}{The total reconnection flux measured from flare ribbon pixels with
brightness at least 4 times the quiescent background for at least 4 minutes; given in the
parenthesis is the difference in the magnetic flux measured in positive and negative magnetic fields, respectively (Section 5.2).}
\tablenotetext{e}{Total heating energy of the flare corona, which is the mean of $\sum\mathcal{E}_{ufc}$
and $\sum\mathcal{E}_{emp}$; the difference between  $\sum\mathcal{E}_{ufc}$ and  $\sum\mathcal{E}_{emp}$
is given in the parenthesis (see Section 5.1).}
\tablenotetext{f}{The maximum coefficient of the time-lagged cross-correlation between two light curves, one being the time derivative
of the GOES SXR 1-8~\AA\ light curve, and the other being the HXR count rates light curve in 12 - 25 keV, or
in 25 - 50 keV by RHESSI, or the total UV 1600~\AA\ counts flux by AIA. A positive time lag indicates that the
time derivative of the SXR light curve lags other light curves. The correlation with HXR light curves is
not available for events \# 3, 7, 11, due to lack of RHESSI observations from the start of the flare.}
\end{deluxetable}
\label{table:1}

\newpage

\end{document}